\documentclass{IEEEtran}
\usepackage[utf8]{inputenc}
\usepackage{cite}
\usepackage{url}
\usepackage{breakurl}
\usepackage{flushend}
\usepackage{bookmark}

\begin{document}
\title{Filtering Video Noise as Audio with Motion Detection to Form a Musical Instrument}
\author{Carl Thomé\\cthome@kth.se}
\maketitle

\begin{abstract}
Even though they differ in the physical domain, digital video and audio share many characteristics. Both are temporal data streams often stored in buffers with 8-bit values. This paper investigates a method for creating harmonic sounds with a video signal as input. A musical instrument is proposed, that utilizes video in both a sound synthesis method, and in a controller interface for selecting musical notes at specific velocities. The resulting instrument was informally determined by the author to sound both pleasant and interesting, but hard to control, and therefore suited for synth pad sounds.
\end{abstract}

\section{Introduction}
In this paper, a musical instrument is proposed that utilizes video in both a sound synthesis method, and in a controller interface for selecting musical notes at specific note velocities.

Light and sound are very different things in the physical domain. However, in the digital domain, as recorded video and audio, the signals share many characteristics. Both are temporal data streams often stored in buffer arrays with 8-bit values.~\cite{dannenberg2003sound} It therefore seems natural to explore ways of producing one with the other, and this paper investigates a specific method for creating harmonic sounds with a video signal as input.

With modern technology having reached a stage where video cameras are everywhere, thanks to the ubiquitous laptop and tablet, it is the perfect time to develop new musical instruments for that particular hardware. Naturally, an ulterior motive exists for this paper: the author wished to design a musical toy that people easily could try out and have fun with, and perhaps be inspired by in their own creative endeavours.

Before reading on about the instrument, trying out an implementation of the proposed instrument is highly recommended, and would likely make the following method description easier to follow. With a laptop fitted with a webcam and a modern web browser, simply visit this link to try out the instrument: \url{www.csc.kth.se/~cthome/motionsynth}

\section{Method}
The proposed instrument consists of three parts: the sound synthesis, the control interface and the resulting visualization which serves as a reinforcement of the control. A basis for the three parts is that they all are directly connected to the video input stream. Audio buffers are created from video frame pixels, the control is basically a motion detector, and the visualization is placed on top of the video stream to highlight detected movements and their corresponding musical tones.

\subsection{Sound synthesis}
The video-to-audio synthesis method used works as follows:
\begin{enumerate}
\item Each video frame is converted to a monochromatic image by calculating the average of the three color channels (red, green and blue) for every pixel.
\item The pixels are converted to an audio buffer by scaling the 8-bit input integer values (0-255) to float values between -1.0 and 1.0.
\item The audio buffer is shuffled to break up contiguous areas of the video frame and to maximize zero crossings in the audio signal.
\end{enumerate}

Thus a solid colored video frame produces silence, and \emph{myrornas krig}~\cite{reddit} (en: static) produces white noise, and the contrast intensity of the noise correlates to the available amplitude in the audio signal.

In order to make the sound harmonic several band-pass filters~\cite{rossing2009science} with extremely narrow peak widths (Q = 1000) were placed across a musical scale, at the corresponding fundamental frequency. Thus each band-pass filter basically corresponds to a note, but additional band-pass filters were also placed at some partials to increase the brightness of the harmonic signal's timbre.

Each group of such band-pass filters (fundamental + some partials) were connected to an Amplitude-Sustain-Release (ASR) envelope~\cite{collins2010introduction}, manipulating both the band-pass filter's amplitude and its Q value. By including the Q value in the envelope the rise and decay of each note duration will be increasingly noisy, so that harmonics fades in and out from noise.

This signal chain from the noise source to the ASR envelope corresponds to a note object representing a musical note that can be played by triggering the ASR envelope. Several such objects in parallel were initialized in the controller, across a musical scale.

\subsection{Control}
The available notes selected were an arbitrary major scale and note objects were initialized across a tessitura equaling the octave span of a piano.

In order to trigger each note, motion detection~\cite{adobe} was employed as follows:
\begin{enumerate}
\item Compare current frame with preceding video frame, and calculate pixel difference.
\item If lots of changed pixels in an area of the video frame, determine the area's 2D position.
\item Use the X-position of the area to select note.
\item Use the Y-position of the area to select the note's velocity.
\end{enumerate}

By moving objects around in front of the camera notes are triggered corresponding to these areas of pixel changes. The motion detection grid size was 51 x 127, corresponding to the note range and velocity range respectively.

\subsection{Visualization}
In order to provide feedback to a performer regarding which notes are currently being played and at what velocity, animated circles per band-pass filter were placed on top of the video signal, at the motion detected areas, but with an animation delay equalling that of the ASR envelope. Thus the performer can see notes rising and decaying in real-time while triggering notes at velocities.

The tessitura was colored in accordance with the light spectrum, with low notes being colored like long wavelengths, and high notes being colored like short wave lengths. The circles were also scaled according to fundamental frequency, as in the physical world where large objects produce lower frequencies when struck, and vice versa.

\section{Results}
An implementation~\cite{motionsynth} of the musical instrument was built in HTML5 with JavaScript, using the Web Audio~\cite{mozilla} and Media Capture and Streams~\cite{adobe} API:s and informally tested by the author. The sound synthesis method was deemed to produce a pleasant harmonic sound, very reminiscent of a typical synth pad sound but with an interesting noise effect at the note attack and decay windows due to band-pass filter peak widths following the ASR envelopes.

The motion detection controller was determined as easy to control, and quite intuitive, but almost impossible to play melodies with. Thus the controller would likely be best served for modulating timbre, or for creating synth pads across a fixed tonal scale.

The instrument seemed to respond well to different environments and was loudest if lighting conditions were varied, and the instrument turns silent if the camera only sees darkness.

\section{Discussion}
Several improvements could be added and some additional exploration should be performed. 
First of all, the instrument should be evaluated formally as well, but due to project time constraints this was cut.

About the instrument itself, the sound could be real-time mastered a little bit with master buss effects such as a reverb convolver to make notes blend together a bit better, or perhaps a ping-pong delay could be employed with the same reasoning. Some tape compression could be used for additional warmth, and so on. However, this might make the coupling between video and audio less evident, even if the sound would become more pleasant.

The controller could use movement velocity as well, and not only perform two-frame motion detection. The velocity of a movement could be used to determine the ASR envelopes attack and release times dynamically.

A similar thought concerns the major scale limitation that was deemed necessary because the controller was not very precise. When a performer moves they tend to move around more than intended so instead of only doing motion detection, perhaps feature detection could be used instead, with finger tip tracking or gesture controls. Then perhaps a chromatic scale would be playable instead.

Furthermore, perhaps the sound synthesis could use all three color channels in parallel, instead of averaging colors, and perhaps specific channels could control different aspects of the sound, such as having a lot of red harshening the timbre by increasing the amount of partials per note. Or if the color balance would weight the band-pass filters maximum amplitude potentials so that the red note in the visualization can only be played if the video contains enough red (preferably on a continuum). With more work, these things could be explored.

Finally, the possibility of a natural coupling between the spatial order of the video data with the produced sound should be explored. By shuffling pixel values and maximizing zero-crossings the signal is noisy enough for the band-pass filters to produce all harmonics, but a fundamental chunk of information in the video data is destroyed in the process. Perhaps it would be more interesting to let the video signal limit the possible notes, or even better: affect the timbre of the sound. Basically, the shuffling stage of the sound synthesis method feels a bit underwhelming, and could be improved upon. The dimension reduction from 2D to 1D should be smarter.

\section{Conclusion}
A method for creating harmonic sounds with a video signal as input has been presented. A musical instrument was proposed, that utilized video in both a sound synthesis method, and in a controller interface for selecting musical notes at specific velocities. The resulting instrument was informally determined by the author to sound both pleasant and interesting, but hard to control, and therefore suited for synth pad sounds. Try it out online now: \url{http://www.csc.kth.se/~cthome/motionsynth/}

\bibliographystyle{ieeetr}
\bibliography{references}

\end{document}